\documentclass[runningheads]{llncs}
\usepackage[utf8x]{inputenc}
\usepackage[T1]{fontenc}
\usepackage{graphicx}
\usepackage{amsmath}
\usepackage{amssymb}
\usepackage{subcaption}
\newtheorem{defi}{Definition}
\begin{document}
\title{Managing Write Access without Token Fees in Leaderless DAG-based Ledgers}
%
%
\author{Darcy Camargo \and
Luigi Vigneri \and
Andrew Cullen}
\authorrunning{D. Camargo et al.}

\institute{IOTA Foundation\\Pappelallee 78/79, 10437 Berlin, Germany\\ \email{name.surname@iota.org}}
\maketitle
\begin{abstract}
A significant portion of research on distributed ledgers has focused on circumventing the limitations of leader-based blockchains mainly in terms of scalability, decentralization and power consumption. Leaderless architectures based on directed acyclic graphs (DAGs) avoid many of these limitations altogether, but their increased flexibility and performance comes at the cost of increased design complexity, so their potential has remained largely unexplored. Management of write access to these ledgers presents a major challenge because ledger updates may be made in parallel, hence transactions cannot simply be serialised and prioritised according to token fees paid to validators. In this work, we propose an access control scheme for leaderless DAG-based ledgers which is based on consuming credits rather than paying fees in the base token. We outline a general model for this new approach and provide some simulation results showing promising performance boosts.

\keywords{Leaderless distributed ledgers \and Dual-token economy \and Priority-based write access \and DAG-based ledgers.}
\end{abstract}
\section{Introduction}
Blockchains have sparked a revolution in the way information is shared in a trustless way. Over the latest decade, research has focused on addressing blockchain's intrinsic shortcomings in search of improved scalability, a more sustainable way of reaching consensus and a fairer distribution of wealth, with the introduction of ``smart transactions''~\cite{wang2019}, new governance solutions and tackling privacy-related issues, among others. One of the main criticisms, though, is still related to performance: Bitcoin and Ethereum, the two most relevant projects by market cap as of 2022, are only able to process a few transactions per second~\cite{bonneau2015}, creating competition between users to obtain writing permission to the blockchain. Such a limited writing space is shared through auction-like mechanisms to discriminate which transactions deserve to be added to the ledger. As transactions compete for the limited writing available, often this system leads to large fees~\cite{roughgarden2017}.

\subsection{Related work}
Various attempts have tried to make DLT projects more scalable, notably lightning networks, sharding and Layer 2 solutions~\cite{zhou2020}. Furthermore, more recently, there has been an increasing interest in directed acyclic graph (DAG) ledgers, which generalize the original chain structure introduced by the blockchain: in fact, when blocks are created at a high rate compared to their propagation time, many competing (or even conflicting) blocks are created leading to frequent bifurcations; DAG-based approaches allow to include blocks not only to the main chain, but also to these bifurcations using additional references~\cite{sompolinsky2016}\cite{mueller2022}. Since transactions can be written and processed in a parallel way, i.e., no total ordering artificially enforcing a pause between subsequent blocks, DAG-based ledgers promise improved throughput and scalability. A number of DAG-based distributed ledger techologies (DLTs) already provide strong performance for consensus and communications layers, such as Honeybadger~\cite{miller2016}, Hashgraph~\cite{baird2020hedera}, Aleph~\cite{Gagol2019} and more recently Narwhal \cite{Danezis2022}. However, the DLTs mentioned still involve leader-based block creation which leaves users exposed to censorship and value extraction by these powerful leaders.

Standard blockchains and leader-based DLTs are built on the dichotomy between the \textit{user} that wants to issue transactions or other state-changing data and a \textit{block issuer} (leader) responsible for creating the blocks that will actually include these data in the ledger. This standard model couples the consensus and access elements of the protocol in the block issuance, creating competition among block issuers to provide ledger access as a service to the base users. In order to have enough incentives for block issuers to invest in this competition, in these protocols users propose fees for their data and the block issuers select which data to include to maximise their profits. Such tokenomics schemes are known for being effective but carry a variety of drawbacks: exclusion of low-scale operations, value extraction from users, fee-bidding wars~\cite{roughgarden2017}, market manipulation, unpredictable pricing and uncertainty of inclusion, to name but a few~\cite{daian2020flash}. 

In order to fulfill the demand of DLT applications that require low to no token fee models, some DLT protocols have attempted to develop zero-fee systems to varied degrees of success \cite{churyumov2016byteball}\cite{lemahieu2018nano}\cite{bagaria2019}\cite{mueller2022}. Among these projects, the DAG-based protocols have shown more promises due to the option of decoupling access and consensus rights, like in Prism~\cite{bagaria2019} or in the IOTA Tangle~\cite{mueller2022}. We adopt such a DAG-based \textit{leaderless} model in this work where users are block issuers. We use this as a basis for developing a novel approach to managing ledger write access that does not require token fees and overcomes many of the negative outcomes of traditional blockchain models.

\subsection{Contributions}
The main contribution of this work is a novel scheme for managing write access to leaderless DAG-based DLTs through \emph{Access Credit}, a quantity that is passively generated based on tokens held and contributions to the protocol (e.g., being a validator). This Access Credit can be \emph{consumed} to create new blocks, buy name services, interact with smart contracts or, in general, use a portion of the DLT resources. The key advantages of the proposed access control scheme are as follows.
\begin{itemize}
    \item \textbf{Zero token fees}: our proposal does not require to pay any tokens to issue blocks; instead, the Access Credit, continuously generated, can be used to create new blocks, whose cost is proportional to its computation and storage requirements as well as the global demand for write access.
    \item \textbf{Leaderless}: contrary to most existing access control schemes for both blockchain and DAG-based architectures, our proposed solution does not rely on leaders or rounds; this greatly improves resilience against censorship and value extraction by powerful block creators.
    \item \textbf{Parallel ledger updates}: with a leaderless  DAG-based ledger we enable parallel execution and writing, as blocks can reference multiple past blocks concurrently.
\end{itemize}

Furthermore, we validate our approach through Python simulations that show the effectiveness of our solution: as we will see, the parallel execution facilitated by the DAG ledger introduces additional complexity to keep ledger consistency across all network participants. We highlight to the reader that we present our proposal in a general manner, that is we do not refer to any existing solution such that the principles described in this paper can be applied to any leaderless DAG-based DLT, and as such, the analysis may lack implementation-specific details. To account for any important omissions, we add an extensive discussion section to present some of the questions that one may encounter while implementing our proposal.

\subsection{Paper structure}
The rest of the paper is organized as follows: the system model is introduced in Section~\ref{sec:system-model}; then, in Section~\ref{sec:access-control}, we present our access control policy; after that, Section~\ref{sec:sim} presents our Python simulations in both single- and multi-node environment. Finally, we present some discussion in Section~\ref{sec:discussion} and conclude our paper in Section~\ref{sec:conclusion}.

\section{System model}\label{sec:system-model}

\subsection{Actors} We categorize the actors of a leaderless DLT as follows.
\begin{itemize}
    \item \textbf{Accounts:} actors capable of holding tokens and, in our proposal, Access Credit. As such, accounts are capable of gaining write access to the ledger by creating blocks. Please note that an account-based DLT is not necessarily mutually exclusive with the UTXO model; in fact, an account can be thought as an identity registered in the ledger to whom one of more UTXOs are associated.
    \item \textbf{Nodes:} the physical machines able to peer with each other to keep local versions of the ledger (related to accounts' token and Access Credit balances) up-to-date through block processing and forwarding.
\end{itemize}

\textbf{Remark:} it is important to note that an account being a block creator in our scheme does not necessarily have the same implications as it does in blockchain architectures. Specifically, block creators do not necessarily act as \emph{validators} do in blockchains, gathering transactions from a shared mempool to include in their blocks. In fact, such a shared mempool is not possible in a DAG-based ledger because blocks are written to the ledger in parallel. Our work focuses on management of write access for accounts, so although these accounts are block issuers, we assume that their motive is to write to their own data to the ledger rather than considering them as intermediaries for base users.

In our model, accounts are the ones including state-changing data into blocks, thus they keep cryptographic signatures to confirm ownership of such data and consume Access Credit to issue the blocks containing such data. \cite{ovezik2022decentralization}. For the sake of completeness, we mention that forms of delegations are possible both for access (through service providers~\cite{cullen2022}) and for consensus (through delegated Proof of Stake~\cite{larimer2014}).

\subsection{Blocks} A block is the fundamental data structure of DLTs carrying value transactions, data, smart contract executions, or any other information that may alter the ledger state. Blocks must also include a cryptographic relation with past blocks, the issuer's signature and fields to manage the consensus protocol (e.g., timestamps). In blockchain technologies, the cryptographic relation is the hash of the block that the issuer believes to be the last included in the ledger. On the other hand, DAG technologies have more malleability: in fact, as multiple blocks may be referenced, the simple act of issuing a block can be used as a statement about trust in numerous blocks. This advantage of DAGs for DLTs was explored in consensus protocols such as \cite{mueller2022}.
As the content of each block may vary greatly, in this paper we define an associated ``cost'', which we call \textit{work}. Work is measured by a protocol gadget as a part of the node software, and it is intended to represent the computational load on the node while processing the block and applying the necessary state changes as well as the resource consumption in terms of bandwidth and storage. As an example, size (in bytes) of the block is one components of the work calculation.

\subsection{Access Credit} The ledger keeps track of computing and storing both Access Credit and token balances. Access Credit is used to gain write access to the ledger. We refer to the amount of spent Access Credits as the \textit{credit consumed}. This quantity needs to be part of the block and it must be signed by the associated account so that the value cannot be altered. The vredit consumed is then used to determine the priority of the block when there is competition to gain write access, as we shall explain in the following section describing our access control. Credits are generated when tokens are moved to a new address through blocks, smart contracts or other means. The amount generated follows the amount of tokens and the time spent in such an address:
\begin{equation}
    \label{CreditGeneration}\text{AccessCredit}  = \text{TokensMoved} \times \text{TimeHeld} - \text{CreditConsumed}.
\end{equation}
When the value in~\eqref{CreditGeneration} is positive, there is a surplus of credits that will be given to a declared account, in an act we call \textit{allotting credits}. Each protocol has its time mechanisms, and the only requirement in the term \textit{TimeHeld} from equation \eqref{CreditGeneration} is that it is objective (so each node agree on how much each account holds of credit). This property is trivial for UTXO-based ledger, but can also be induced in any other protocol by using appropriate timestamping mechanisms.

\section{Credit-based access control}\label{sec:access-control}

In this section, we present the Credit-based access control mechanism for leaderless DLTs using DAG ledgers for parallel writing and execution. We stress that our proposal does \textit{not} make use of token fees.

\subsection{Block creation}

As we mentioned, accounts are the actors that include the state-changing data (e.g., value transactions, smart contracts) in the blocks. 
They are required to interact with a node to set a reasonable amount of Access Credit consumed and to forward the block to the rest of the DLT network. In fact, nodes can be thought as gateways to the network and play a fundamental role in the congestion management of the entire architecture: accounts (that can be managed through wallets or light nodes) do not receive nor process the blocks produced in the network. Consequently, they ignore the current congestion level and are enable to properly set the amount of Access Credit to consume. Hence, accounts must either set up their own nodes or use third-party free or paid services (e.g., through access service providers).

Upon request, nodes send information related to the real-time congestion level of the newtork, namely an estimation of the amount of Access Credit needed to successfully schedule a block. Then, account can take a more or less conservative approach when setting the credit consumed of the newly created block depending on the account's preferences, similarly to the way priority is set when gas fees are paid in Ethereum~\cite{leonardos2021}. In this work, we assume that consumed Access Credit is a quantity larger or equal to $0$: while bounds are useful for spam protection (in the case of a lower bound) and to avoid overspending resources (in the case of an upper bound), we leave the study of this optimization as a future work.

\subsection{Access control}

In this section, we describe our proposed access control mechanism for leaderless DAG-based DLTs. Unlike standard blockchains where block producers try to extract value by selecting the most profitable blocks, in our approach the rules are defined at protocol level and each node participates without the possibility of extracting value. Our access control chooses which blocks get gossiped in the peer-to-peer network, where the Access Credit is consumed instead of being redistributed, hence nodes have no incentives to deviate from the protocol.

In the following, we present the main components of our proposal, namely the enqueueing phase, the scheduling mechanism and the policy to drop blocks during congestion.

\subsubsection{Enqueuing}
As blocks get gossiped, receiving nodes verify the correctness of their content (verification procedure varies depending on the specific implementation). For valid blocks, the protocol calculates the \textit{Priority Score}, defined as follows.
\begin{defi}[Priority Score]
    Consider a block $B$ and the tuple $(c_B,w_B)$ where $c_B$ is the Access Credit consumed and $w_B$ the work of block $B$. We refer to Priority Score $S_B$ the ratio between the Access Credit consumed and the work of block $B$:
\begin{equation}
    S_B = \frac{c_B}{w_B}.
\end{equation}
\end{defi}

Once the Priority Score is computed, the block is enqueued into the \textit{scheduling buffer}, which gathers all blocks not yet scheduled (more details in the Scheduling subsection). In our proposal, this buffer is a \textit{priority queue} sorted by Priority Score. The insert of a new block in the buffer has linear complexity.

\subsubsection{Scheduling}
The scheduling policy is a mechanism that selects which blocks must be forwarded in the DLT network. We consider a scheduler that works in a service loop, where every $\tau$ units of time it selects the blocks with the largest Priority Score in the scheduling buffer such that the work units of the selected blocks are smaller or equal than $m$ work units. In this scenario, the enforced network throughput limit is $m/\tau$ work units per second.

When a block is chosen to be scheduled, it is forwarded to neighbouring nodes where it can be enqueued in their buffer if they have not yet received it, after which the block undergoes the same scheduling process in each new node. We do not assume any specific gossip protocol: flooding, i.e., forwarding indiscriminately to all neighbours, is a popular choice in DLTs, but this can be optimised to save network bandwidth.

\subsubsection{Block drop}
In practice, scheduling buffers have a limited size. In fact, the usage of large buffers in networks has been proved to be detrimental to performance~\cite{gettys2011}. In this work, we use a simple policy to drop blocks when the buffers get full, namely the protocol will drop the block with the lowest priority score, removing it from the buffer. Additionally, to limit the effectiveness of long-range attacks, we also drop blocks whose timestamps become older than a certain threshold compared to the node's local clock. 

\textbf{Remark:} when blocks do not get finalized, i.e., they do not receive enough references, the Access Credit consumed is ``reimbursed'' to the issuer's account. The reimbursement should happen when the consensus mechanism reaches finalization on the state of non-inclusion of the data in the ledger state. The exact details on how long the data is kept and the reimbursement takes is specific to each consensus mechanism, and thus protocol, being out of the scope of the write access mechanism of this paper.



\section{Simulations}\label{sec:sim}

This section shows a performance analysis concerning the credit-based access control proposed in Section~\ref{sec:access-control}. We first introduce the simulation setup in Section~\ref{sec:sim-setup}. Then, in Section~\ref{sec:single-node-sim}, we analyse the performance of the access control by looking at \textit{a single node}: this allows to collect metrics related to cost of new blocks, time spent in the scheduling buffer, scheduled and not scheduled blocks per account, etc. Finally, in Section~\ref{sec:multi-node-sim}, we present the outcomes of experiments on a \textit{multi-node setting} to verify ledger consistency and analyse the rate of discarded blocks.
 
\subsection{Simulation setup}\label{sec:sim-setup}

\begin{figure}
    \centering
    \includegraphics[width=0.75\textwidth]{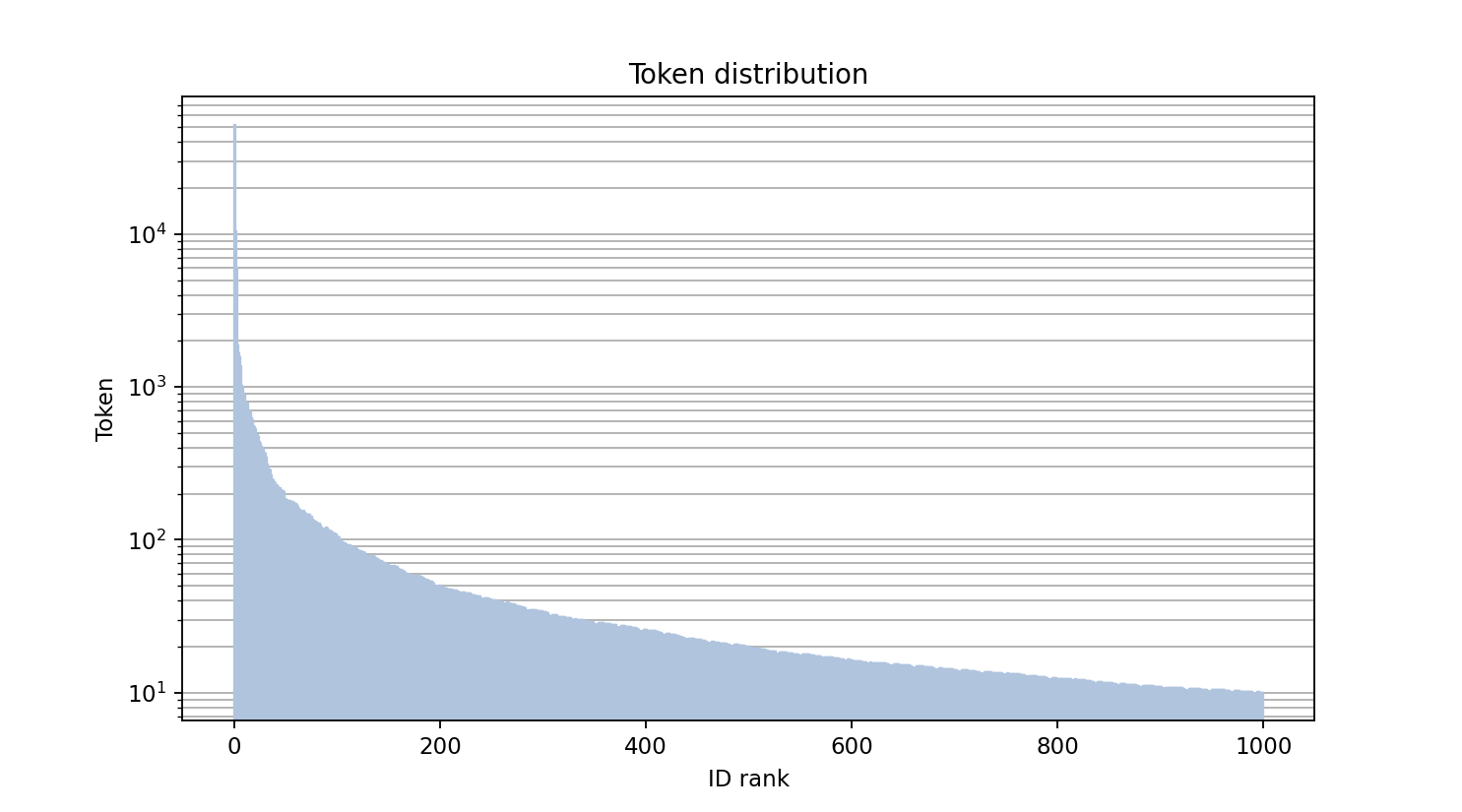}
    \caption{Token held by account.}
    \label{fig:tokens}
\end{figure}

In our setup, we consider $1000$ accounts, i.e., block issuers. The token holdings belonging to those issuers are drawn from a Power Law distribution of the form
\begin{equation}
    p(x) = \frac{\alpha-1}{x_{min}}\cdot\left(\frac{x}{x_{min}}\right)^{-\alpha},
\end{equation}
where $\alpha = 2$ and minimum token $x_{min} = 10$. A visual representation of the token holdings sorted by tokens can be found in Figure~\ref{fig:tokens}. The amount of tokens per issuer does not vary over the course of the simulations. Furthermore, each user gets $1$ credit/second for every $10$ tokens held: for example, a user with $25$ tokens obtains $2.5$ credit/second.\newline

Blocks are generated according to a non-homogenerous Poisson Process, with alternating congested and uncongested periods of 3 minutes each. We define a congested period as the time interval where total block generation rate\footnote{This is the sum of the block generation rate over all accounts.} is larger than scheduling rate. The simulation is run for one hour, that is $10$ congested periods and $10$ uncongested periods. Additionally, we impose a scheduling rate of $100$ blocks per second, i.e., a block is scheduled every $10$ ms (for the sake of simplicity, all blocks have the same size). In our simulations, the number of blocks issued by an account is proportional to its token holdings. Moreover, we define four types of block issuers according to the way the block cost is set\footnote{We stress that more sophisticated strategies are expected to optimize the performance of the system and will be studied as a future work.}:
\begin{itemize}
    \item \textbf{Impatient:} These accounts consume all of their Access Credits each time they issue a block, so their credit consumption per transaction is high when they do not have many transactions to issue, and the credit consumption per transaction is low when they have a large number of transactions to issue. They do not respond in any way to the credit consumption they see in the buffer.
    \item \textbf{Greedy:} These accounts look at the highest amount of Access Credit consumed in the scheduling buffer and consume $1$ more Access credit than this. If this greedy policy dictates that they would need to consume more than they have, they simply do not issue anything until the price goes down or they have generated enough Access Credit.
    \item \textbf{Gambler:} These nodes consume the amount of Access Credit of one of the top 20 blocks in the priority queue, chosen randomly.
    \item \textbf{Opportunistic:} These nodes consume $0$ Access Credit, regardless of what is seen in the scheduling buffer. Traffic from these nodes is perfectly acceptable during periods of low congestion, but constitutes spam during congested periods and it is expected to be dropped from scheduling buffers.
\end{itemize}

Finally, we assume the buffer having a maximum capacity of $500$ blocks. Blocks are removed from the buffer when one of the two scenario happens:
\begin{itemize}
    \item \textbf{Full buffer:} If the buffer contains $500$ blocks, a newly arrived block can be added to the buffer if and only if it consumes more Access Credits than those consumed by at least one block in the buffer; in this case, the latter will be removed and replaced with the newly arrived block.
    \item \textbf{Maximum time in the buffer:} When a block spends $30$ seconds in the buffer without being scheduled, it gets immediately removed.
\end{itemize}

Potential changes in the parameters used in the multi-node simulator will be explicitly mentioned in Section~\ref{sec:multi-node-descr}.


\subsection{Single-node simulator}\label{sec:single-node-sim}

\begin{figure}
\centering
    \begin{subfigure}[b]{0.32\textwidth}
        \includegraphics[width=\textwidth]{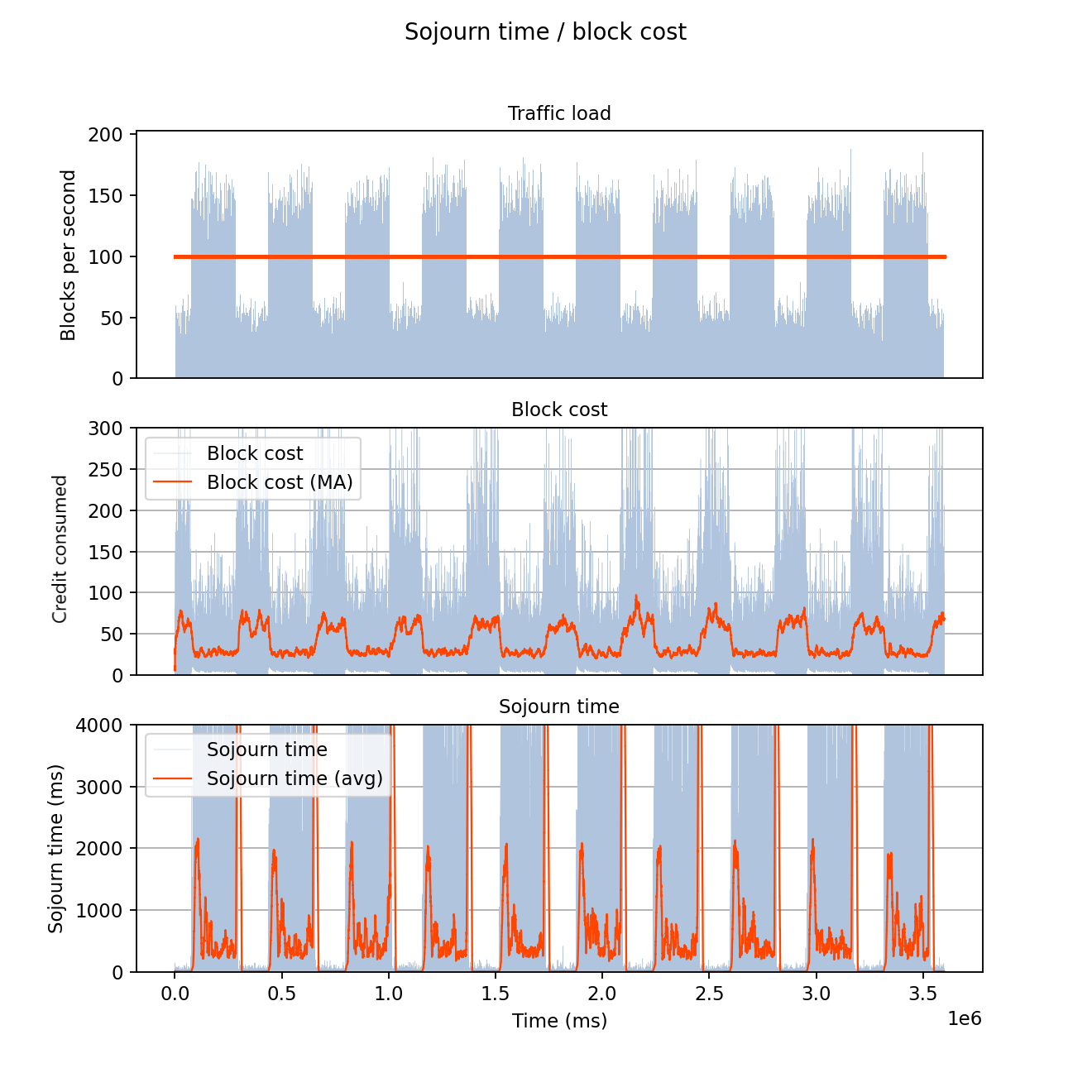}
        \caption{All accounts are \textit{impatient} users.}
        \label{fig:soj-cost-anxious}
    \end{subfigure}
    \begin{subfigure}[b]{0.32\textwidth}
        \includegraphics[width=\textwidth]{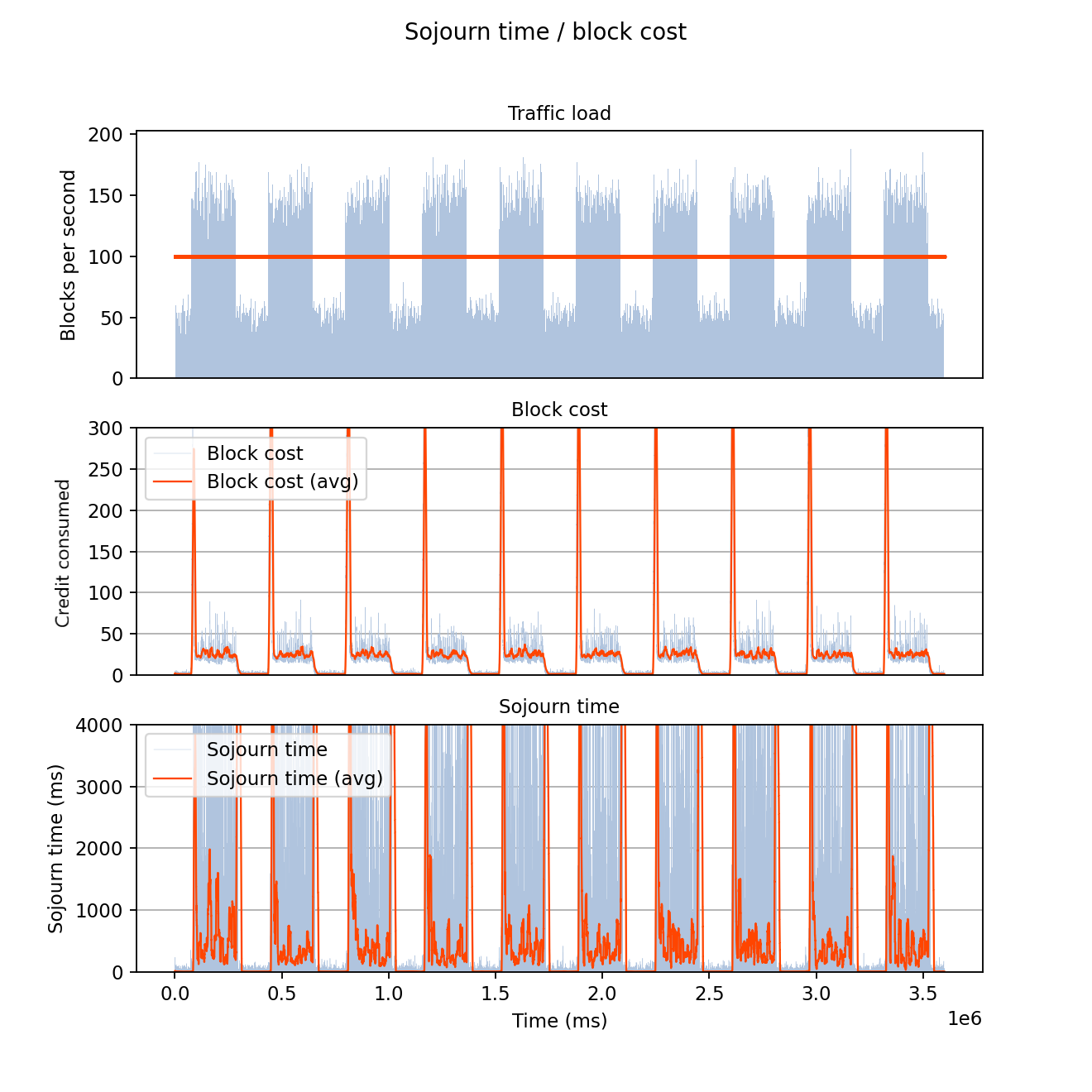}
        \caption{All accounts are \textit{greedy} users.}
        \label{fig:soj-cost-greedy}
    \end{subfigure}
    \begin{subfigure}[b]{0.32\textwidth}
        \includegraphics[width=\textwidth]{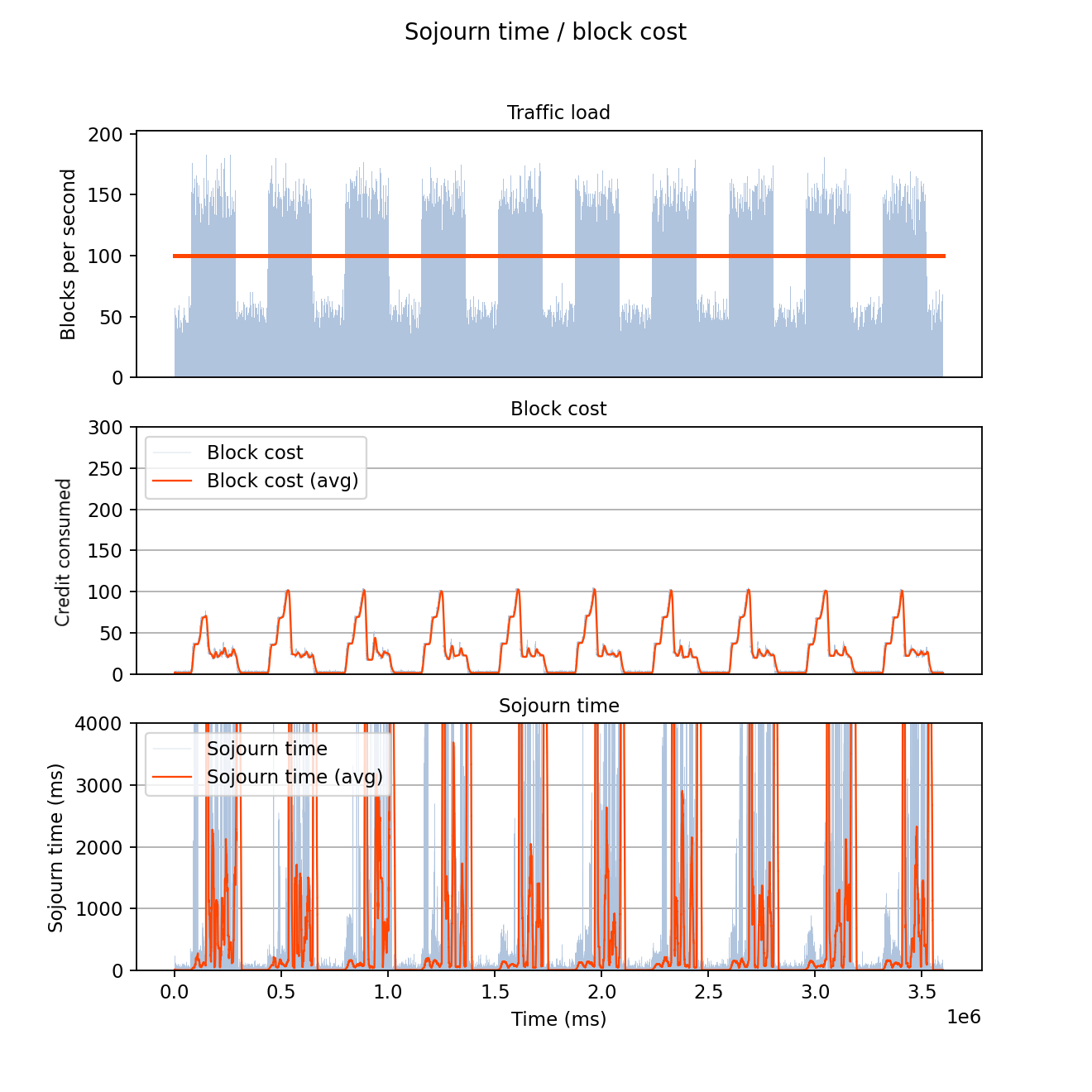}
        \caption{All accounts are \textit{gambler} users.}
        \label{fig:soj-cost-gambler}
    \end{subfigure}
    \caption{Traffic load (top figure), credits consumed (middle) and sojourn time (bottom) per block over time. Red line indicates the scheduling rate in the traffic load plot, and the moving average in the other plots.}
\end{figure}

\subsubsection{Impatient strategy}

In this set of simulations, all accounts follow the \textit{impatient} consumption strategy. In Figure~\ref{fig:soj-cost-anxious}, we plot the cost of a scheduled block and the sojourn time of the same block as the simulation time advances. As a reference, we also add the traffic load over time, which alternates congested and uncongested periods. When accounts act as \textit{impatient} users, we realize that the cost of a block increases during less congested periods, while -- during congestion -- the cost of a block stabilizes at less than $30$ credits with peaks up to $150$ credits; conversely, in uncongested periods the credits spent is at least the double. This is because users tend to overspend when using this consumption strategy: during congestion, accounts have less time to accumulate Access Credit; the plot basically shows how much one can accumulate since the latest block it has issued -- accumulation is larger if blocks are issued less often.

The sojourn time, defined as the time a block spends in the scheduling buffer (remember, this is a single-node simulator, so the sojourn time is the time spent in a single buffer), is very low when the network is uncongested but experiences large oscillations during congestion: in particular, after the transition to congestion, the mean sojourn time spikes to around $2$ seconds and then keeps oscillating between $0.5$ and $1$ second; a non-negligible number of blocks experience a much larger sojourn time as it can be seen by the blue line in Figure~\ref{fig:soj-cost-anxious}.

\subsubsection{Greedy strategy}

Here, we show the same set of plots, but when all accounts act as \textit{greedy}. A greedy consumption strategy seems to optimize the inefficiencies of the impatient one, which tends to overspend unnecessarily. The cost of a block, from Figure~\ref{fig:soj-cost-greedy}, is now very low (close to 0) with little traffic; however, the transition to a congested network creates a very steep increase in the cost of a scheduled block: for a short period of time, the average consumed Access Credits is larger than $300$, then suddenly decreasing around $30$. This strategy can be compared with first price auctions, carrying their intrinsic drawbacks as well. While several recent approaches have been trying to mitigate the fluctuations in the block cost and to improve the user experience~\cite{roughgarden2017}\cite{leonardos2021}, we stress that finding an optimal credit consumption policy is out of the scope of this paper.

Similarly, it is possible to see frequent oscillations in the sojourn times with spikes (i) at the beginning and (ii) at the end of the congested period: (i) the increased traffic load alters the dynamics of the system, lowering the rank in the priority queue for blocks not yet scheduled, and we notice that oscillations are visible throughout the entire congested period; (ii) additionally, when congestion ends, a lot of blocks sitting in the buffer for long (but not yet dropped) have the opportunity to be scheduled experiencing a large delay, witnessed by the spike at the end of each high-traffic period.

\subsubsection{Gambler strategy}

In this set of simulations we have all accounts with the \textit{gambler} strategy. There are clear differences with the previous scenarios: in Figure~\ref{fig:soj-cost-gambler}, we see that the spikes in the Credits consumed are largely reduced compared to the \textit{greedy} scenario: we cannot see accounts consuming more than $100$ credits. However, the cost of scheduled blocks stabilizes at a price only marginally lower than the previous scenarios.

\subsubsection{Mixed strategy}

In this scenario, we allow users with different consumption strategies to coexist. Specifically, $10\%$ of accounts is \textit{impatient}, $60\%$ is \textit{greedy} and the rest is \textit{gambler}. This set of simulations aim to provide a more realistic environment where multiple types of users share the network.

\begin{figure}
\centering
    \includegraphics[width=0.75\textwidth]{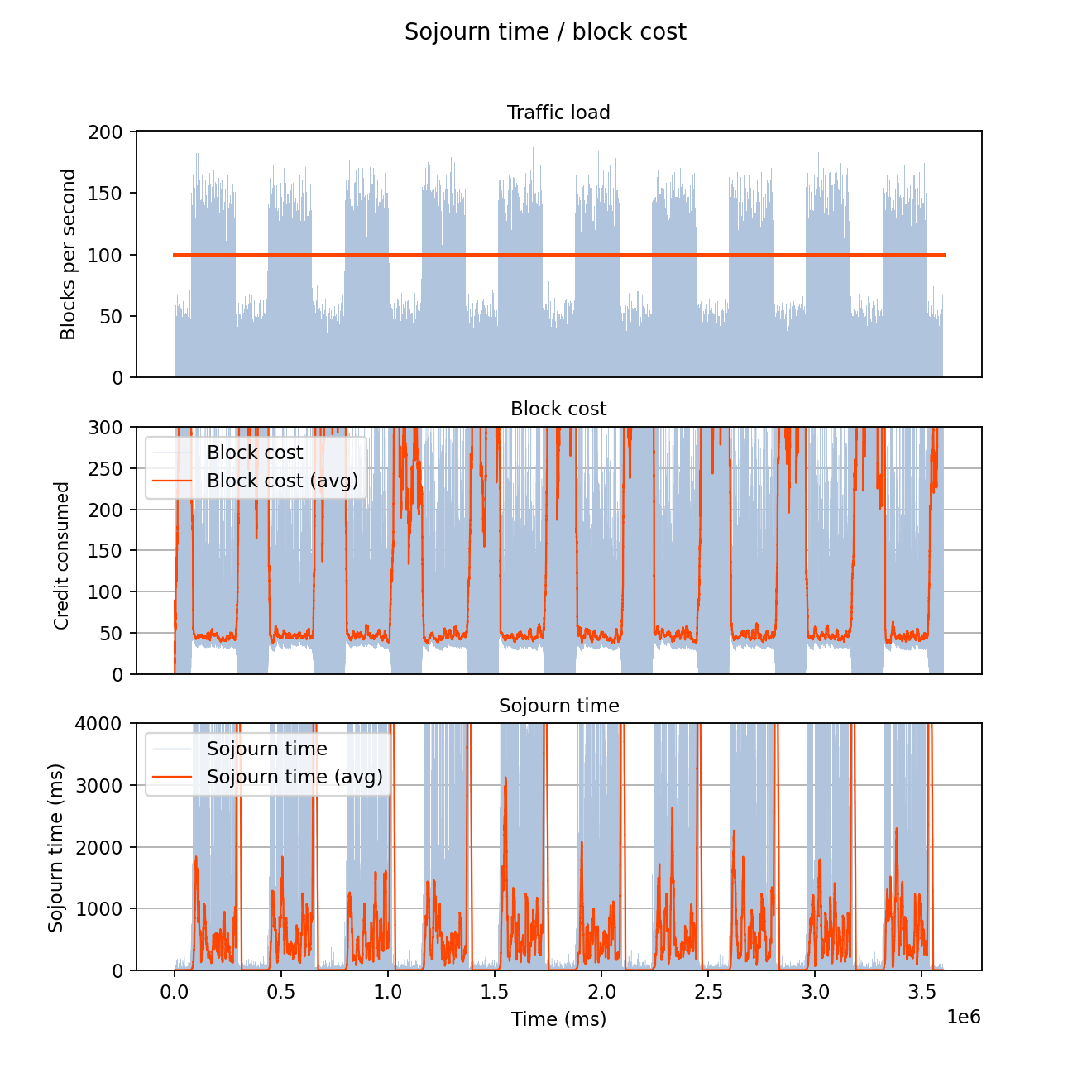}
    \caption{Traffic load (top figure), credits consumed (middle) and sojourn time (bottom) per block over time in the mixed scenario. Red line indicates the scheduling rate in the traffic load plot, and the moving average in the other plots.}
    \label{fig:soj-cost-mixed}
\end{figure}

In Figure~\ref{fig:soj-cost-mixed} we see that the average cost of a scheduled block is still driven large by impatient accounts, although such nodes represent only the $10\%$ of the total block issuers. Similar to previous scenarios, we also see large oscillations in the sojourn times during congestion with peaks up to $2$ seconds.

\begin{figure}
    \centering
    \includegraphics[width=0.75\textwidth]{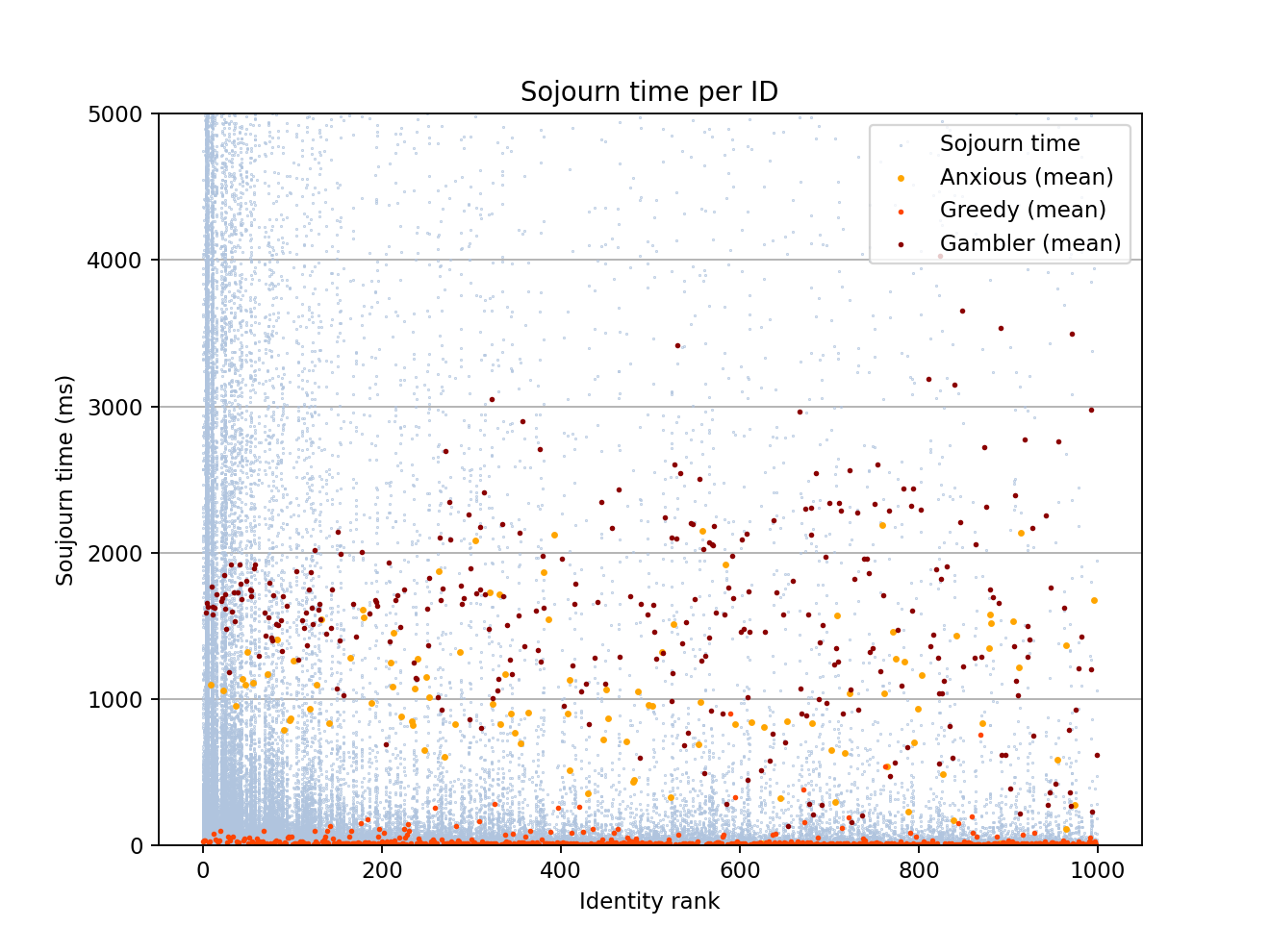}
    \caption{Sojourn time and related mean per account.}
    \label{fig:soj-mixed}
\end{figure}

An interesting consideration can be done with respect to Figure~\ref{fig:soj-mixed}, which decouples the sojourn time per account differentiating between impatient (yellow), greedy (red) and gambler (dark red): we observe that the mean sojourn time for greedy issuers is much lower than the other policies. While a large sojourn time is expected for \textit{gambler}, it should not be the case for \textit{impatient}. The explanation is that greedy users do \textit{not} issue blocks if they do not have enough Access Credits: basically, these accounts have a self-regulating \textit{rate setter}, and the benefits in terms of improved delays are clearly visible.

\subsection{Multi-node simulator}\label{sec:multi-node-sim}

\subsubsection{Description of the simulator}\label{sec:multi-node-descr}
The following simulations are implemented in a multi-node simulator which emulates a complete DAG-based DLT protocol, i.e., each node maintains a copy of the DAG, uses a selection algorithm to choose where attach new blocks and checks the validity of all arriving blocks. A number of specific DAG-based protocol details are included which our proposal do not necessarily depend on, but this allows us to at least provide prelimenary results for integrating this approach into a working protocol. Each node also operates an account for issuing blocks, so we use the terms node and account interchangeably in this section. The same consumption policies are tested as for the single-node simulator, but we use a smaller network and shorter simulation times to facilitate detailed presentation of each node's outcome.

\begin{figure}
\centering
\includegraphics[width=0.75\textwidth]{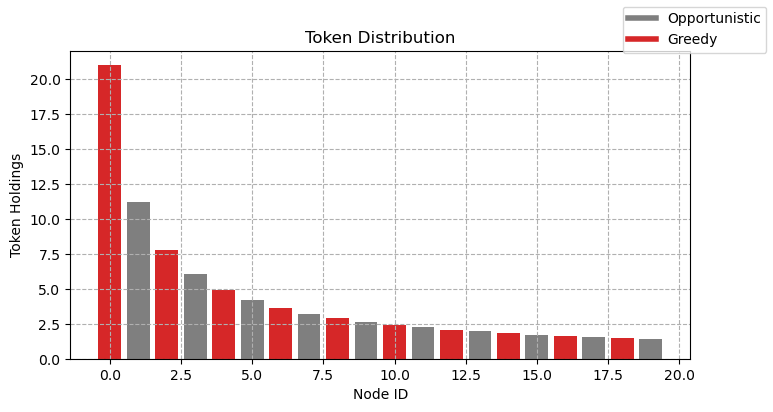}
\caption{Token distribution and credit consumption policies for 20 nodes.}
\label{fig: mb tokens 1}
\end{figure}

The simulations consist of 20 nodes connected in a random 4-regular graph topology i.e., 4 neighbours each. The communication delays between nodes are uniformly distributed between 50 ms and 150 ms. The scheduling rate is 25 blocks per second. We use the same token distribution as in the single-node simulator. We initially consider a mix of greedy and opportunistic nodes with token holding distribution as illustrated in Figure~\ref{fig: mb tokens 1}.

We slightly modify the block generation process in this set of simulations: here, blocks are generated according to a separate Poisson process for each node and added to the node's local mempool from which they can create blocks. For the first minute of each simulation, blocks are generated at 50\% of the scheduling rate, then for the following two minutes, the rate gets to 150\% compared to scheduling rate, and for the final minute, it decreases to the 50\% again. This traffic pattern simply seeks to show one cycle of increase in demand and then subsiding of demand.

Finally, we introduce the concept of block \textit{confirmation} in the DAG through Cumulative Weight (CW):
\begin{defi}[Block confirmation]
    The $CW_B\in\mathbb{N}^+$ of block $B$ indicates how many times $B$ has been referenced directly or indirectly by other blocks. If $CW_B \geq 100$, then a node locally considers block $B$ as confirmed.
\end{defi}
\begin{defi}[Confirmation Rate]
A block is confirmed when all nodes have marked the block as confirmed. Confirmation rate is the rate at which blocks become confirmed.
\end{defi}
CW in a DAG is analogous to the depth of a block in a blockchain which is often used for confirmation. Additionally:
\begin{defi}[Dissemination Rate]
A block is disseminated when all nodes have seen the block. Dissemination rate is the rate at which blocks become disseminated.
\end{defi}

\textbf{Remark:} ``Scaled'' plots are scaled by the node's ``fair share'' of the scheduler throughput which is proportional to their token holdings, so a scaled rate of 1 means they are getting 100\% of their fair share. In plots showing metrics for all nodes, the thickness of the trace corresponding to each node is proportional to the token holdings of that node.



\subsubsection{Experimental results}
\begin{figure}
\centering
\includegraphics[width=0.75\textwidth]{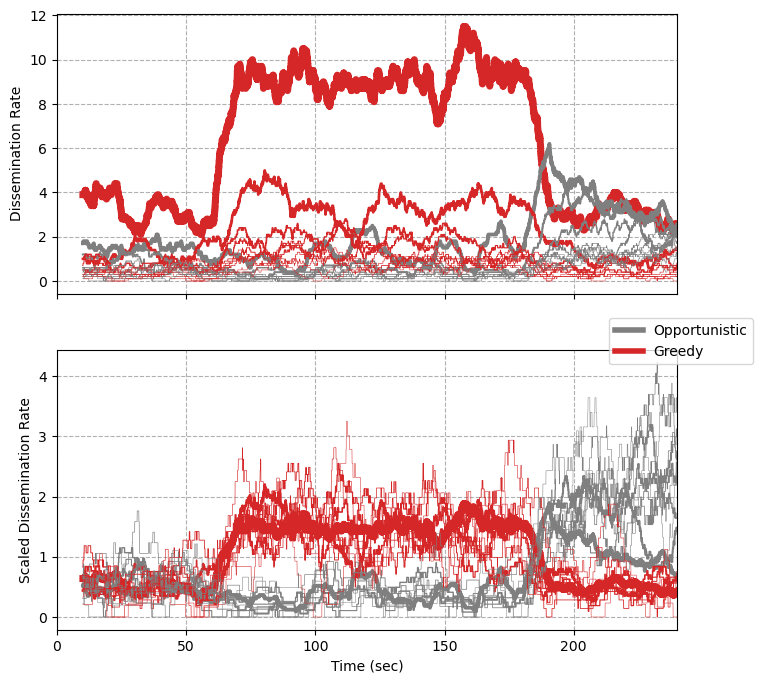}
\caption{Dissemination rates and scaled dissemination rates. The scaled rate shows the rate relative to the account's token holdings.}
\label{fig: mb dissemrates}
\end{figure}

\begin{figure}
\centering
\includegraphics[width=0.75\textwidth]{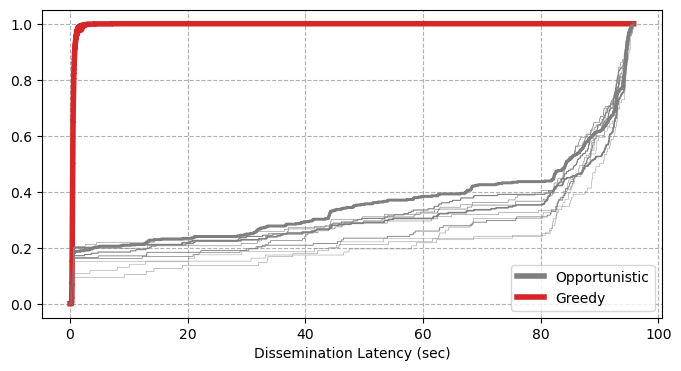}
\caption{Cumulative density function of the dissemination latency.}
\label{fig: mb dissemlatency}
\end{figure}

We begin by considering the dissemination rates, as seen in Figure~\ref{fig: mb dissemrates}. Here, the greedy nodes are able to issue more than their ``fair share'' because the opportunistic nodes are opting not to consume any Credits and hence the greedy nodes get priority from the scheduler by consuming more. Figure~\ref{fig: mb dissemlatency} displays the corresponding dissemination latency of each node's blocks as a cumulative density. This paints a similar picture, with greedy nodes experiencing lower delays than their opportunistic counterparts.

\begin{figure}
\centering
\includegraphics[width=0.75\textwidth]{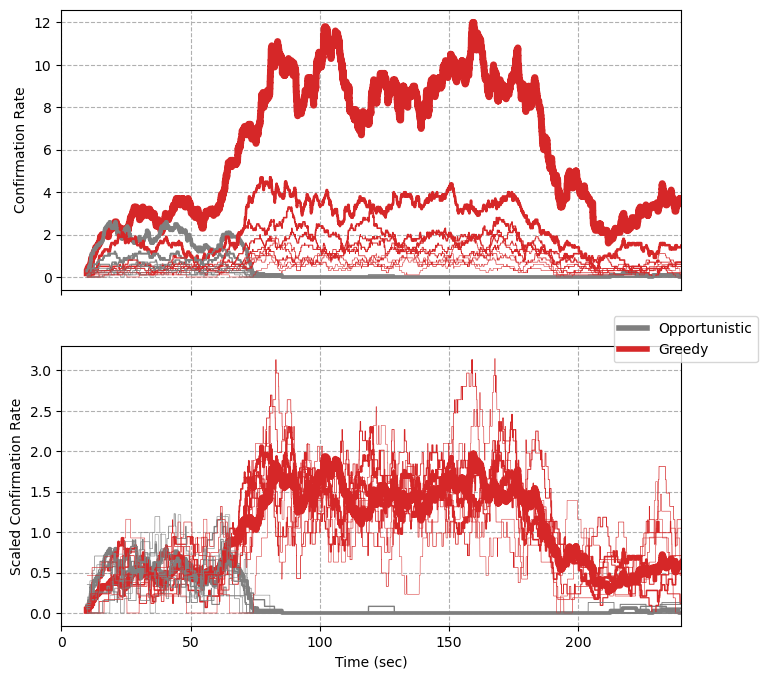}
\caption{Confirmation rates and scaled confirmation rates. The scaled rate shows the rate relative to the account's token holdings.}
\label{fig: mb confrates 2}
\end{figure}

Figure~\ref{fig: mb confrates 2} illustrates the confirmation rates corresponding to this simulation. These traces follow a very similar trajectory to the dissemination rates, but we notice that even when congestion dies down, the confirmation rates of the opportunistic nodes do not recover immediately as the dissemination rates did. This is due to the fact that many old delayed blocks from the congested period are stuck in the buffers of nodes across the network and as these old blocks begin to be forwarded when the congestion goes away, they are not selected by other nodes to attach to, so their cumulative weight does not grow and they do not become confirmed.

These multi-node simulator results only present a very limited scenario with basic credit consumption policies, but the results show promise for providing effective access control. However, they also begin to show some of the complexities of integrating this approach into complete DAG-based DLT protocols. Further studies need to be carried out for specific DAG implementations to fully understand the implications of our approach.

\section{Discussion}\label{sec:discussion}

\subsection{Economic Incentives}

As we discussed before, fees can be used to regulate access to DLT, but can also bring detrimental properties, such as the possibility of extracting value from users and the creation of inconsistencies in access. Nevertheless, fees provide essential incentives for many protocols, usually being the way security is ensured. In order to create sustainable economic incentives, we expect that Access Credits will have an active market, where users can sell their spare access. This creates a positive feedback loop where the generated gains received are in form of access which further incentives network adoption and usage. 


\subsection{Limiting the accumulation of Access Credit}\label{sec: bounding credits}

One can notice that the way Access Credit is defined makes this quantity highly inflationary, as credits are passively generated by tokens even when they are not in use. This could lead to situations where congestion leads to an amount of Access Credits consumed being excessively high, which would make access prohibitive for some periods of time, or to situations where accounts can accumulate enough Access Credits to continuously spam the network for some time. To counteract these events, we propose to introduce a concave function $F(t)$ increasing with time, where $F(0)=0$, such that Eq.~\eqref{CreditGeneration} becomes:
\begin{align*}
    \text {AccessCredit}  = \ &\text{TokensMoved} \times F(\text {TimeHeld})\\
    &-\text{CreditConsumed}.
\end{align*}

This slows down the accumulation and, depending on the function chosen can even cap it, e.g., when $F(t) \propto (1-e^{-\gamma t})$. Using a concave function in the $\textit{TimeHeld}$ factor has a side effect: it pushes one to create blocks to allot credits more often, since Access Credit generation is faster soon after tokens are moved. The Access Credits being consumed to issue block will work as an offset to this: if an account allots credits from their tokens $n$ times during the $\mbox{TimeHeld}$ interval and consuming the same amount of Access Credits each time, the balance by the end would be
 \begin{align*}
      \mbox{AccessCredit}_n = \ & \text {TokensMoved} \times nF(\mbox{TimeHeld}/n)\\
      &-n\text{CreditConsumed}.
 \end{align*}
Hence, the Access Credit generated over this generated will have a maximal value in $n$. 

\subsection{The negative Access Credit problem}
 Due to the distributed nature of DLTs, Access Credit balances may become temporarily negative either due to natural network delay or due to malicious behavior (similar to nothing-at-stake problem). In this paper, we have not touched yet the scenario where accounts reach negative balance on Access Credits. The most effective solution is to process all blocks from the account with negative balance, find consensus on which ones should be accepted and punish the offending account after this process.
 
\textbf{Remark:} it is not possible to  filter out blocks leading to negative Access Credit balances to avoid forks in nodes' local views of the DAG. Suppose a malicious account sends two blocks, $A$ and $B$, where processing only one of them would not cause its balance to go negative but processing both would. A subset of the nodes in the network may process $A$ and filter $B$, while the other subset may process $B$ and filter out $A$. This would create a problematic scenario where nodes will have inconsistent views of the ledger. An attacker could then repeat this procedure with many blocks, creating many possible forks. 

\section{Conclusion}\label{sec:conclusion}

We have proposed a credit-based access control mechanism for leaderless DAG-based DLTs. Our solution solves the problem of regulating write access to DAG-based DLTs without the need for token fees or serialisation of ledger updates into blocks by validators. The proposal is based on \emph{Access Credits}, which are naturally generated for holding the base token. State-changing data must consume these credits to be included in the ledger, creating a utility loop where rewards are given in Access Credits.

Our simulations show that under varied user behaviors, the consumed credits remain stable over time, even with large jumps of demand for ledger write access. Additionally, we showed that write access can be effectively regulated across multiple nodes in a peer-to-peer network in some simple scenarios.
Leaderless DAG-based ledgers present enormous potential for advances in the DLT field, and this work will provide a foundation for similar schemes seeking to manage write access in these systems in the future. 

\bibliographystyle{IEEEtran}
\bibliography{bibliography}

\end{document}